\documentclass[aps,prd,twocolumn,amsmath,amssymb,superscriptaddress,nofootinbib,showpacs,preprintnumbers]{revtex4-1}
\usepackage{aas_macros}
\usepackage{graphicx}
\usepackage[colorlinks=true]{hyperref}
\usepackage[usenames]{color}

\begin{document}

\preprint{CERN-TH-2018-214,  KCL-PH-TH/2018-55}
\title{Scattering of light dark matter in atomic clocks}

\author{Peter Wolf}\email{Peter.Wolf@obspm.fr}
\affiliation{SYRTE, Observatoire de Paris, Universit\'e PSL, CNRS, Sorbonne Universit\'e, LNE, 75014 Paris, France}

\author{Rodrigo Alonso}
\email{Rodrigo.Alonso@cern.ch}
\affiliation{Theoretical Physics Department, CERN, CH-1211 Geneva 23,
 Switzerland}

\author{Diego Blas}
\email{Diego.Blas@cern.ch}
\affiliation{Theoretical Physics Department, CERN, CH-1211 Geneva 23,
 Switzerland}
\affiliation{Theoretical Particle Physics and Cosmology Group, Department of Physics, King's College London, London WC2R 2LS, UK}

\date{\today}


\begin{abstract}
We present a detailed analysis of the effect of light Dark Matter (DM) on atomic clocks, for the case where DM mass and density are such that occupation numbers are low and DM must be considered as particles scattering off the atoms, rather than a classical field. We show that the resulting atomic clock frequency shifts are first order in the scattering amplitudes, and particularly suited to constrain DM models in the regime where the DM mass $m_\chi \ll$ GeV. We provide some rough order of magnitude estimates of sensitivity that can be confronted to any DM model that allows for non zero differential scattering amplitudes of the two atomic states involved in the clock.
\end{abstract}

\maketitle

\section{Introduction}

In recent years there has been increasing interest, both theoretical and experimental, in the possibility of ultra light (typically $<10^{-10}$~eV) dark matter (DM) detection using atomic clocks \cite{Derevianko2014,Arvanitaki2015,Stadnik2015a,VanTilburg2015,Hees2016,Wcislo2016a, Roberts2017a,Hees2018}. Most of that work is using a DM model where the DM is a massive scalar field that is non-minimally coupled to standard matter, with that field being described classically either as coherent oscillations or as some topological defects. The outstanding stability and accuracy of cold atom clocks has been used to set some of the most stringent limits on the coupling constants in this type of DM models \cite{VanTilburg2015,Hees2016,Wcislo2016a, Roberts2017a}.

In this work, and in more detail in the companion paper \cite{Alonso2017}, we consider more general DM models (beyond only a classical scalar field) that may include non-scalar DM, which interacts with the spin of the atoms in atomic clocks. Additionally for high masses, or if not all of the DM density in the galaxy is of this type, the mass and density may be such that occupation numbers are $\leq 1$, so that a classical field description is no longer appropriate. Then DM must be treated as particles that scatter of the atoms in atomic clocks. This opens up a new range of laboratory DM searches in mass/density regions that cannot be probed by the usual WIMP or Axion experiments, whilst being different from the ultra low mass scalar fields described above.

More particularly we elaborate the theory for the scattering of some particle with mass $m_\chi$ off the atoms with mass $m_a$ for the particular case where $m_\chi\ll m_a$ (the mass range where $m_\chi$ is close to $m_a$ is already well explored). We explicitly calculate the frequency and phase shift of atomic clocks as a function of the scattering amplitudes for different types of clocks (e.g. Ramsey and Rabi spectroscopy) and give expressions that allow confronting these results to any model for such particles and their interaction with the fundamental particles of the standard model. This analysis is quite general and applies to any light particle scattering off atoms, DM or not.

We then provide some very rough benchmark values for the sensitivities that could be achieved in terms of the corresponding scattering amplitudes in different scenarios that could apply to DM. In a companion article \cite{Alonso2017} we then use those results to estimate the sensitivities in terms of the parameters (masses, coupling constants) of some particular DM models. But, the present article is kept more general, the aim being to provide rough numbers that allow evaluating the potential sensitivity of atomic techniques to any DM model, provided it allows the derivation of scattering amplitudes for collisions of the light DM particles with atoms.

The scattering of particles off cold atoms in clocks has been discussed most recently in \cite{Gibble2013} in the context of frequency shifts in atomic clocks related to the presence of residual background gases. One of the results of that analysis is that in cold atom clocks only forward scattered atoms can reach the detection region, which leads to a loss of Ramsey fringe amplitude related to the interference between the scattered and unscattered wave in the forward direction. As we show below, this is strictly speaking only true for the case where the initial momentum of the scatterers $m_\chi v_{\chi i}$ is much larger than the maximum change of momentum $m_a \Delta v_{a}$ of the clock atoms that still allows them to reach the detection region. However, in this work we are interested in the opposite situation where the scatterers are very light so that $m_\chi v_{\chi i} \ll m_a \Delta v_{a}$, and the conclusions in that case are phenomenologically different. For example, and as one would expect, there is now no loss of fringe amplitude as, to leading order in $m_\chi/m_a$, the cold atoms do not change their trajectory in the scattering process, and thus all of them make it to the detection region. Furthermore, we extend our analysis to Rabi spectroscopy rather than only Ramsey spectroscopy as in \cite{Gibble2013}, which turns out to be a promising methods for actually observing such collisions.
 
Our article is organised as follows: We first recall some basics of quantum scattering theory in section \ref{sec:GenTheory} with particular emphasis on light scatterers in section \ref{sec:IIB}. We then apply it to atomic clocks in Ramsey (sect. \ref{sec:Ramsey}) or Rabi (sect. \ref{sec:Rabi}) configuration. The final section \ref{sec:DM_strategies} discusses some strategies for DM detection using the previously obtained results, and provides some very rough sensitivity estimates based on present day technology.

\section{Quantum scattering of two particles}

The theory presented in this section is well known and can be found in many textbooks (e.g. \cite{Gottfried2003}) or lecture courses (e.g. \cite{Hitoshi}). Its application to cold atom clocks has been discussed most recently in \cite{Gibble2013}.

\subsection{General theory} \label{sec:GenTheory}

Consider a clock atom $a$ interacting with some background particle $\chi$. The two particle wave function is $\Psi(t,{\bf x}_a,{\bf x}_\chi)$. The probability of detecting the atom in the detection region $D$ at time $t_D$ is
\begin{equation}\label{equ:PD}
P_D = \int_D d^3x_a \int_{-\infty}^\infty d^3x_\chi |\Psi(t_D,{\bf x}_a,{\bf x}_\chi)|^2.
\end{equation}
The evolution of $\Psi(t,{\bf x}_a,{\bf x}_\chi)$ is governed by the two particle Schr\"odinger equation where the Hamiltonian includes the interaction potential $V({\bf x}_a-{\bf x}_\chi)$. With the change of variables to the centre of mass and relative positions ${\bf x}_G = (m_a{\bf x}_a+m_\chi {\bf x}_\chi)/(m_a + m_\chi)$ and ${\bf x} = {\bf x}_a-{\bf x}_\chi$ the two particle wavefunction can be factorised to $\Psi(t,{\bf x}_a,{\bf x}_\chi) = \Psi_G(t,{\bf x}_G)\Psi(t,{\bf x}))$ where $\Psi_G(t,{\bf x}_G)$ is the centre of mass wavefunction and $\Psi(t,{\bf x})$ that of the ``virtual particle" with mass $\mu=m_\chi m_a/(m_a+m_\chi)$ (see e.g. \cite{Cohen-Tannoudji1977}). The latter is obtained from the one particle Schr\"odinger equation in relative coordinates ${\bf x} = {\bf x}_a-{\bf x}_\chi$, with the reduced mass $\mu$ and the interaction potential $V({\bf x})$. 
 
Let us first consider the case where $m_a \ll m_\chi$. Then, to zeroth order in $m_a/m_\chi$ the centre of mass position is that of the scatterer ${\bf x}_G \simeq {\bf x}_\chi$ and the detection probability (\ref{equ:PD}) becomes
\begin{eqnarray}\label{equ:PD2a}
P_D &\simeq& \int_D d^3x |\Psi(t_D,{\bf x})|^2 \int_{\infty} d^3x_\chi|\Psi_G(t_D,{\bf x}_\chi)|^2 \nonumber \\
&=& \int_D d^3x |\Psi(t_D,{\bf x})|^2.
\end{eqnarray}
This case is treated in \cite{Gibble2013}. It corresponds to considering only forward scattering of the virtual particle (essentially the atom in this case) as the integral is carried out only over the detection region, and cold atoms scattered away from that region are lost. Phenomenologically one observes a frequency shift and a loss of fringe amplitude, the latter corresponding to cold atoms being scattered away from the detection region.

The opposite case $m_a\gg m_\chi$ is the main subject of our article. We now have to zeroth order in $m_\chi/m_a$: ${\bf x}_G \simeq {\bf x}_a$ and the detection probability (\ref{equ:PD}) becomes
\begin{eqnarray}\label{equ:PD2}
P_D &\simeq& \int_{\infty} d^3x |\Psi(t_D,{\bf x})|^2 \int_D d^3x_a|\Psi_G(t_D,{\bf x}_a)|^2 \nonumber \\
&=& \int_{\infty} d^3x |\Psi(t_D,{\bf x})|^2,
\end{eqnarray}
as $\Psi_G(t_D,{\bf x}_a) \sim \Psi_a(t_D,{\bf x}_a)$ which now propagates entirely to the detection region as the centre of mass coincides with the atom and propagates freely along its initial direction. Phenomenologically there is no loss of fringe amplitude now (all atoms make it to the detection region) but a frequency shift remains, as we will see below. Note that this applies entirely to cold atoms, provided that they are more massive than the scatterers, and in particular that $m_\chi v_{\chi i} \ll m_a \Delta v_{a}$ as mentioned in the introduction. Equation (\ref{equ:PD2}) simply translates the fact that any, i.e. not only forward, scattering of the virtual particle (essentially the $\chi$ particle in this case) still allows the atom to be detected.

To make the criterion for using (\ref{equ:PD2a}) or (\ref{equ:PD2}) more physical for atomic clocks we still assume that $m_a > m_\chi$ but work to first order in $m_\chi/m_a$. Classical energy-momentum conservation shows that the scattering angle $\theta$ of the virtual particle (i.e. the change in direction of ${\bf v} = {\bf v}_a-{\bf v}_\chi$) is given by

\begin{equation}
{\rm tan}\theta \simeq \mp \frac{m_a \Delta v_{a}}{m_\chi v_{\chi i}} \frac{\left(1-\frac{m_a^2}{4 m_\chi^2}\frac{\Delta v_{a}^2}{v_{\chi i}^2}\right)^{1/2}}{\left(1-\frac{m_a^2}{2m_\chi^2}\frac{\Delta v_{a}^2}{v_{\chi i}^2}\right)}. \label{B10}
\end{equation}
where $v_{\chi i}$ is the initial velocity of the light particle in the rest frame of the clock atom, and $\Delta v_{a}$ is the maximum allowed change of velocity of the clock atom that still allows it to reach the detection region.

As an order of magnitude consider Cs clock atoms that are scattered off thermal H$_2$ molecules ($v_{\chi i} \approx 10^3$~m/s), as in \cite{Gibble2013}. For typical Cs fountain clock parameters the clock atoms are still detected for $\Delta v_a \approx 10^{-3}$~m/s, corresponding to a 1~cm detection aperture 0.5~s after the scattering event. Then the terms in brackets in (\ref{B10}) are $\simeq 1$ and we have $\theta < 10^{-3}$~rad, which is quite compatible with considering only forward scattering (equivalent of (\ref{equ:PD2a})) as done in \citep{Gibble2013}. On the contrary for DM particles with e.g. $m_\chi \approx 1$~keV and $v_{\chi i} \approx 10^{-3} c$ the maximum possible value of $\Delta v_a$ is $\approx 10^{-3}$~m/s and thus the atom is detected for any value of $\theta$,  which yields equation (\ref{equ:PD2}).

\subsection{Particular case of light scatterers} \label{sec:IIB}

We now further investigate the case where $m_a\gg m_\chi$. We want to evaluate the integral (\ref{equ:PD2}) with $\Psi(t_D,{\bf x})$ the wavefunction of the virtual particle, solution of the single particle Schr\"odinger equation with mass $\mu$ and position ${\bf x} = {\bf x}_a-{\bf x}_\chi$. Assuming a short range potential $V({\bf x})$ that solution is standard in scattering theory \cite{Gottfried2003, Hitoshi} and takes the general form $\Psi_{out} = \Psi_{inc} + \Psi_{sc}$, where $\Psi_{out}$ is the outgoing wavefunction after scattering,
\begin{equation} \label{equ:psi_inc}
\Psi_{inc} = N e^{i(kz-\omega t)} e^{-\frac{x^2+y^2+(z-vt)^2}{4d^2}},
\end{equation}
is the incident wave with $N = \left(\frac{1}{2\pi d^2}\right)^{3/4}$, $\omega = \frac{\hbar k^2}{2 \mu}$, $v = \hbar k/\mu$, and $d$ the width of the Gaussian wave packet. The scattered wave,
\begin{equation} \label{equ:psi_sca}
\Psi_{sc} = N \frac{f(\theta)}{r} e^{i(kr-\omega t)} e^{-\frac{(r-vt)^2}{4d^2}},
\end{equation}
is a spherical wave moving outwards with Gaussian profile in $r$ and scattering amplitude $f(\theta)$, where $\theta$ is the scattering angle.

The detection probability is then given by (\ref{equ:PD2}) in the form 
\begin{eqnarray} \label{equ:det_int}
P_D &=& \int_\infty d^3x |\Psi_{out}|^2  \nonumber \\
&=& \int_\infty d^3x\left(|\Psi_{inc}|^2 + |\Psi_{sc}|^2 + |\Psi_{int}|^2 \right),
\end{eqnarray}
with the interference term $|\Psi_{int}|^2 \equiv \Psi_{inc}^*\Psi_{sc} + \Psi_{inc}\Psi_{sc}^*$.

The integral of the first term in (\ref{equ:det_int}) is equal to one.
The second term is simply given by probability conservation which requires that $\int_\infty d^3x (|\Psi_{sc}|^2 + |\Psi_{int}|^2) =0$.

The third term can be calculated explicitly \cite{Gottfried2003, Hitoshi} giving
\begin{equation} \label{final_int}
\int_\infty d^3x|\Psi_{int}|^2  = -\frac{1}{2\pi d^2} \frac{4 \pi}{k} {\rm Im}[f(0)].
\end{equation}
This term represents the ``fractional shadow" cast by the scatterer on the detection region i.e. the part of the incident wave `lost" into the scattered one, and is known as the "optical theorem". Note that only the forward scattering amplitude ($\theta = 0$) is involved, as the interference term quickly averages to zero for $\theta \neq 0$.\footnote{In the case where $m_\chi v_{\chi i} \gg m_a \Delta v_{a}$, (\ref{final_int}) describes the loss of cold atoms, and thus fringe amplitude, due to scattering  \cite{Gibble2013}.}

Obviously, one then has a detection probability in (\ref{equ:det_int}) that is always $P_D=1$ and unaffected by the scattering. However, as we will show below, for an initial superposition state the scattering may lead to a phase shift between the two parts of the superposition that can be measured.

\section{Application to atomic clocks}\label{sec:main}

\subsection{Ramsey spectroscopy} \label{sec:Ramsey}

The detailed theory of atomic clocks and Ramsey spectroscopy can be found in e.g. \cite{Vanier1989a}. Here we consider a simplified situation with two ideal $\pi/2$ pulses separated by a free evolution time $T$, and calculate the frequency shift from a single scattering event taking place during time $T$ after the first and before the second pulse.

The ground and exited internal states of the atom are $|1\rangle$ and $|2\rangle$. The initial state before the first $\pi/2$ pulse is
\begin{eqnarray}\label{equ:pulse1}
\langle 2|\Psi\rangle &=& 0, \nonumber \\
\langle 1|\Psi\rangle &=& \Psi_{inc} ,
\end{eqnarray}
with $\Psi_{inc}$ given by (\ref{equ:psi_inc}). After the first pulse the state is
\begin{eqnarray}
\langle 2|\Psi\rangle &=& \frac{1}{\sqrt{2}}(-i \Psi_{inc}), \nonumber \\
\langle 1|\Psi\rangle &=& \frac{1}{\sqrt{2}}\Psi_{inc} ,\\
\nonumber
\end{eqnarray}
where we have set the phase of the light to zero. The scattering takes place between the two pulses, and at the second pulse the light has phase $\phi$. The state after the second pulse is then
\begin{eqnarray} \label{equ:final_state}
\langle 2|\Psi\rangle &=& \frac{1}{2}(-i \Psi^{(2)}_{out} - i \Psi^{(1)}_{out}e^{-i\phi}), \nonumber \\
\langle 1|\Psi\rangle &=& \frac{1}{2}(\Psi^{(1)}_{out} - \Psi^{(2)}_{out}e^{i\phi}) ,
\end{eqnarray}
where, as before, $\Psi_{out} = \Psi_{inc} + \Psi_{sc}$, and the superscript refers to scattering in state $|1\rangle$ or $|2\rangle$ i.e. we assume different scattering amplitudes $f_1(\theta)$ or $f_2(\theta)$ in (\ref{equ:psi_sca}). The probability of detecting e.g. the internal state $|2\rangle$ is then
\begin{widetext}
\begin{eqnarray}
P_2 &=& \int_\infty d^3x|\langle 2|\Psi\rangle|^2  = \frac{1}{4}\int_\infty d^3x|\Psi_{inc}(1+e^{-i\phi})+\Psi^{(2)}_{sc} + \Psi^{(1)}_{sc}e^{-i\phi}|^2  \nonumber\\
&=& \frac{1}{2}(1+{\rm cos}\phi) + \frac{1}{4}\int_\infty d^3x\left(|\Psi^{(2)}_{sc}|^2 + |\Psi^{(1)}_{sc}|^2 + |\Psi^{(2)}_{int}|^2 + |\Psi^{(1)}_{int}|^2 \right. \nonumber \\
&+& \Psi^{(2)}_{sc} \Psi^{(1)*}_{sc}e^{i\phi} + \Psi^{(2)*}_{sc} \Psi^{(1)}_{sc}e^{-i\phi} + \left. \Psi_{inc}^*\Psi^{(2)}_{sc} e^{i\phi}+\Psi_{inc}\Psi^{(2)*}_{sc} e^{-i\phi} + \Psi_{inc}^*\Psi^{(1)}_{sc} e^{-i\phi}+\Psi_{inc}\Psi^{(1)*}_{sc} e^{i\phi}\right) . \label{P2_0} 
\end{eqnarray}
\end{widetext}
Probability conservation (see section \ref{sec:IIB}) requires that the first four terms in the integral of (\ref{P2_0}) cancel pairwise. To evaluate the remaining six terms we first re-write $\Psi^{(\gamma)}_{sc} \equiv f_\gamma(\theta) \Psi_0$ where $\gamma=1,2$ and the explicit form of $\Psi_0$ can be simply read off (\ref{equ:psi_sca}). With that notation (\ref{P2_0}) becomes
\begin{widetext}
\begin{eqnarray} \label{P2_1}
P_2 &=& \frac{1}{2}(1+{\rm cos}\phi) + \frac{1}{4}\int_\infty d^3x\left(f_1^* f_2 e^{i\phi} |\Psi_0|^2 + f_1 f_2^* e^{-i\phi} |\Psi_0|^2 \right. \nonumber \\
&+& \left. f_2 e^{i\phi}\Psi_{inc}^*\Psi_0 + f_2^* e^{-i\phi} \Psi_{inc}\Psi_0^*  + f_1 e^{-i\phi}\Psi_{inc}^*\Psi_0 + f_1^* e^{i\phi} \Psi_{inc}\Psi_0^*\right). 
\end{eqnarray}
\end{widetext}
For the first two terms in the integral we evaluate the $r$ integral with $|\Psi_0|^2$ from (\ref{equ:psi_sca}), and for the last four terms we apply the optical theorem (\ref{final_int}) to obtain
\begin{eqnarray} \label{P2_2}
P_2 &=& \frac{1}{2}(1+{\rm cos}\phi) 
+ \frac{1}{4 \pi d^2}\left({\rm Re}[e^{-i \phi}F_{12}]\right) \nonumber \\
&-& \frac{1}{2 k d^2}\left({\rm Im}[e^{-i \phi}(f_1 - f_2^*)]\right),
\end{eqnarray}
where we have defined $F_{12} \equiv \int d\Omega f_1(\theta) f_2^*(\theta)$ and with the scattering amplitudes in the last line of (\ref{P2_2}) in the forward direction only i.e. $f_\gamma=f_\gamma(0)$. More explicitly,
\begin{eqnarray} \label{P2_3}
P_2 &=& \frac{1}{2}(1+{\rm cos}\phi) \nonumber \\
&+& \frac{1}{d^2}\left(\frac{1}{4\pi}{\rm Re}[F_{12}]-\frac{1}{2k}{\rm Im}[f_1 + f_2]\right){\rm cos}\phi \nonumber \\
&+& \frac{1}{d^2}\left(\frac{1}{4\pi}{\rm Im}[F_{12}]+\frac{1}{2k}{\rm Re}[f_1 - f_2]\right){\rm sin}\phi.
\end{eqnarray}

For $f_1 = f_2$ we have ${\rm Im}[F_{12}]=0$. The optical theorem together with (\ref{equ:psi_sca}) gives directly ${\rm Re}[F_{12}] = 4\pi {\rm Im}[f]/k$. So equ. (\ref{P2_3}) reduces to its standard form in the absence of scattering, as expected.

The quantity $F_{12}$, and more generally (\ref{P2_3}), can be further expanded in terms of partial wave phases (see Appendix \ref{app:partial}). However, for small cross-sections we will be interested in the region where $k |f| \ll 1$, which implies that, of the scattering terms in (\ref{P2_3}), the last one will dominate by a factor $\simeq 1/(k|f|)$.\footnote{Unless $|f_1-f_2| \ll |f|$, but we will not consider that case as it provides much lower sensitivity.} The result is thus simplified to 
\begin{eqnarray} \label{P2_4}
P_2 &\simeq & \frac{1}{2}(1+{\rm cos}\phi) + \frac{1}{2kd^2}{\rm Re}[f_1 - f_2]{\rm sin}\phi.
\end{eqnarray}

A similar calculation for state $|1\rangle$ yields
\begin{eqnarray} \label{P1_3}
P_1 &\simeq & \frac{1}{2}(1-{\rm cos}\phi) - \frac{1}{2kd^2}{\rm Re}[f_1 - f_2]{\rm sin}\phi.
\end{eqnarray}
The sum satisfies $P_1+P_2=1$ for any value of $\phi, f_1, f_2$, as expected when there is no loss of atoms due to scattering, i.e. in our case where $m_\chi/m_a \ll 1$.

To calculate the frequency shift we set $\phi = \delta T$ in (\ref{P2_4}), where $\delta \equiv \omega-\omega_{12}$ is the detuning of the light from atomic resonance. We then expand for $\delta T \ll 1$ and find the value $\delta_{max}$ for which $P_2$ is maximum, i.e. the position of the central Ramsey fringe maximum, by setting $\partial P_2/\partial \delta =0$. The result is
\begin{equation} \label{d_max_Ramsey2}
\delta_{max} \simeq \frac{1}{kd^2T}{\rm Re}[f_1(0)-f_2(0)],
\end{equation}
where we recall that in (\ref{P2_4}) $f_\gamma = f_\gamma(0)$.

\subsubsection*{Discussion}
The change of detection probability as given by equation (\ref{P2_4}) corresponds to the frequency  in (\ref{d_max_Ramsey2}), but to no loss of fringe amplitude.

The physical interpretation is relatively straightforward: the scattering event corresponds to the atom experiencing the short-range interaction potential $V({\bf x})$ for a short time. If that potential is different for the two atomic states, then the atomic resonance is shifted by that difference (the two states are ``dressed" differently by the potential). Therefore the scattering event leads to a transient shift of the atomic transition frequency $\omega_{12}$, thus to a small change of the phase $\phi$. During the Ramsey time $T$ this phase shift leads to a shift of the resonance frequency of $\delta_{max}$ as given in (\ref{d_max_Ramsey2}).

The process described above is independent of which direction the atoms or $\chi$ particles are scattered into, but only the atoms that arrive at the detection region contribute to the signal. In our case of $m_a \gg m_\chi$ all atoms make it to the detection region and there is no loss of signal amplitude. In the opposite case $m_a \ll m_\chi$ (or more precisely $m_\chi v_{\chi i} \gg m_a \Delta v_{a}$, see section \ref{sec:GenTheory}) only forward scattered atoms make it to the detection region and contribute to the frequency shift, so there is a shift {\it and} a signal loss \cite{Gibble2013}. 

In any case, the effect on the atomic transition frequency is also present at zero momentum transfer, which makes it ideal to look for very light scatterers or coherent effects as compared to other techniques based on atomic recoil.

\subsection{Rabi spectroscopy} \label{sec:Rabi}

The physical interpretation from the previous section suggests that the effect of scattering on the clock frequency might be different for different types of interrogation sequences of the atom. Consider, for example, a single scattering event. In the case of a Ramsey sequence of duration $T$ the result on the detection probability is (\ref{P2_4}) irrespective of when the scattering takes place, as a transient change of $\omega_{12}$ has the same effect whether it happens toward the beginning, middle or end of $T$. However for a single Rabi pulse of duration $T$ one would expect the effect to be maximum for the scattering occurring at $T/2$ and minimum for scattering near the beginning or the end of the pulse. This can be easily understood using the image of fictitious spin on a Bloch sphere or the sensitivity function formalism described e.g. in \cite{Santarelli1998}. An explicit derivation is provided below.

Consider a Rabi interrogation with a single pulse of duration $T$ during which the scattering takes place. The initial state of the atom at $t=0$ is 
\begin{eqnarray}\label{equ:initial}
\langle 2|\Psi\rangle &=& 0, \nonumber \\
\langle 1|\Psi\rangle &=& \Psi_{inc}.
\end{eqnarray}

The collision takes place at time $t=t_c$. Just before the collision the state is (see e.g. \cite{Cohen-Tannoudji1977,Vanier1989a,Young1997})
\begin{eqnarray} \label{equ:Rabi_before}
\langle 2|\Psi\rangle &=& -i e^{-i\delta t_c/2}{\rm sin}\theta \ {\rm sin}\left(\frac{\Omega_r t_c}{2}\right) \Psi_{inc} \\
\langle 1|\Psi\rangle &=& e^{i\delta t_c/2}\left[{\rm cos}\left(\frac{\Omega_r t_c}{2}\right)+i {\rm cos}\theta \ {\rm sin}\left(\frac{\Omega_r t_c}{2}\right) \right]\Psi_{inc}, \nonumber 
\end{eqnarray}
where $\delta$ is the detuning, $\Omega$ the Rabi frequency, $\Omega_r = \sqrt{|\Omega|^2+\delta^2}$ is the off-resonant Rabi frequency, ${\rm sin}\theta \equiv \Omega/\Omega_r$, and ${\rm cos}\theta \equiv -\delta/\Omega_r$, and where we have set the initial phase of the light to zero for simplicity.

The state just after the collision is then
\begin{eqnarray}
\langle 2|\Psi\rangle &=& -i e^{-i\delta t_c/2}{\rm sin}\theta \ {\rm sin}\left(\frac{\Omega_r t_c}{2}\right) \Psi_{out}^{(2)},  \\
\langle 1|\Psi\rangle &=& e^{i\delta t_c/2}\left[{\rm cos}\left(\frac{\Omega_r t_c}{2}\right)+i {\rm cos}\theta \ {\rm sin}\left(\frac{\Omega_r t_c}{2}\right) \right]\Psi_{out}^{(1)}, \nonumber
\end{eqnarray}
and at time $T$
\begin{widetext}
\begin{eqnarray} \label{equ:Rabi_T}
\langle 2|\Psi\rangle &=& e^{-i\delta T/2}\left\lbrace-i {\rm sin}\theta \ S_c \left[C_T-i{\rm cos}\theta \ S_T \right] \Psi_{out}^{(2)} \right. 
- \left. i {\rm sin}\theta \ S_T \left[C_c+i {\rm cos}\theta \ S_c \right]\Psi_{out}^{(1)} \right\rbrace, \nonumber \\
\langle 1|\Psi\rangle &=& e^{i\delta T/2}\left\lbrace-{\rm sin}^2\theta \ S_T S_c \Psi_{out}^{(2)} \right.
+ \left. \left[C_T+i{\rm cos}\theta \ S_T\right] \left[C_c+i {\rm cos}\theta \ S_c \right] \Psi_{out}^{(1)} \right\rbrace,
\end{eqnarray}
\end{widetext}
where we have used the shorthand notation $S_c \equiv {\rm sin}(\Omega_r t_c/2)$, $C_c \equiv {\rm cos}(\Omega_r t_c/2)$, $S_T \equiv {\rm sin}(\Omega_r (T-t_c)/2)$, and $C_T \equiv {\rm cos}(\Omega_r (T-t_c)/2)$. It is easy to check that, as required, (\ref{equ:Rabi_T}) reduces to the standard expression (i.e. (\ref{equ:Rabi_before}) with $t_c=T$) when $t_c=0$ and setting $\Psi_{out}^{(1)} = \Psi_{out}^{(2)} = \Psi_{inc}$ i.e. in the absence of scattering.

The probability of detecting the atomic state 2 is obtained by integrating over all space. Thus
\begin{widetext}
\begin{eqnarray}
P_2 &=& \int_\infty d^3x|\langle 2|\Psi\rangle|^2   \nonumber \\
&=& {\rm sin}^2\theta \ \int_\infty d^3x \left[S_c^2 |X_T|^2\left(|\Psi_{inc}|^2 + |\Psi^{(2)}_{sc}|^2 + |\Psi^{(2)}_{int}|^2 \right) \right.+ S_T^2 |X_c|^2\left(|\Psi_{inc}|^2 + |\Psi^{(1)}_{sc}|^2 + |\Psi^{(1)}_{int}|^2 \right) \nonumber \\
&&~~~+ \left. S_cS_T\left((X_T^*X_c+X_TX_c^*)|\Psi_{inc}|^2 + X_TX_c^*\Psi^{(2)}_{sc} \Psi^{(1)*}_{sc} + X_T^*X_c\Psi^{(2)*}_{sc} \Psi^{(1)}_{sc} + |\overline{\Psi}^{(2)}_{int}|^2 + |\overline{\Psi}^{(1)}_{int}|^2\right) \right], \label{equ:P2_Rabi_1}
\end{eqnarray}
\end{widetext}
where $|\overline{\Psi}_{int}^{(2)}|^2 \equiv X_TX_c^*\Psi_{inc}^*\Psi_{sc}^{(2)} + X_T^*X_c\Psi_{inc}\Psi_{sc}^{(2)*}$, $|\overline{\Psi}_{int}^{(1)}|^2 \equiv X_T^*X_c\Psi_{inc}^*\Psi_{sc}^{(1)} + X_TX_c^*\Psi_{inc}\Psi_{sc}^{(1)*}$  and we have used another shorthand notations $X_T \equiv C_T-i{\rm cos}\theta \ S_T$ and $X_c \equiv C_c+i{\rm cos}\theta \ S_c$.

As in the Ramsey case, probability conservation requires that the last two terms in the first and second line of (\ref{equ:P2_Rabi_1}) cancel. The last four terms of the third line are identical to the last six terms of (\ref{P2_0}) with the substitutions $\phi \rightarrow \Delta\alpha$ and $1/4 \rightarrow S_cS_TA_cA_T$, where we have defined $X_T \equiv A_Te^{i\alpha_T}$, $X_c \equiv A_ce^{i\alpha_c}$, and $\Delta\alpha \equiv \alpha_T-\alpha_c$.   

The result is then similar to (\ref{P2_4}),
\begin{eqnarray} \label{equ:P2_Rabi_2}
P_2 &=& {\rm sin}^2\theta \ \left\lbrace {\rm sin}^2\left(\frac{\Omega_r T}{2}\right)\right.  \\
&+& \left. S_cA_TS_TA_c\frac{1}{2\pi d^2}\frac{4\pi}{k}{\rm Re}[f_1 - f_2]{\rm sin}(\Delta\alpha)\right\rbrace , \nonumber
\end{eqnarray}
the second line of which is obviously equal to zero for $f_1=f_2$ as required. Similarly, as required, one recovers the standard result when $t_c=0$ or $t_c=T$ as in that case $S_c=0$ or $S_T=0$ and the scattering term vanishes.

Expression (\ref{equ:P2_Rabi_2}) can be simplified when choosing $T$ such that $\Omega_rT = \pi$, as in standard Rabi spectroscopy. We then expand for small $\delta$ i.e. $\delta/\Omega \ll 1$. In the expansion we keep terms up to ${\cal O}(\delta^2)$ in the first (classical) term of (\ref{equ:P2_Rabi_2}) and up to ${\cal O}(\delta)$ in the second (scattering) term, leading to:
\begin{equation} \label{P2_Rabi_5}
P_2  \simeq  1-\frac{\delta^2}{\Omega^2} +\frac{2 S_c^2C_c^2}{kd^2}{\rm Re}[f_1-f_2]\left(\frac{2}{{\rm sin}(\Omega t_c)}\frac{\delta}{\Omega}\right), 
\end{equation}
where at the required order $\Omega \simeq \Omega_r$.

Setting the derivative $\partial P_2/\partial \delta =0$ we find the detuning for which $P_2$ is maximum, i.e. the frequency shift:
\begin{equation} \label{d_max_Rabi2}
\delta_{max}(t_c) \simeq {\rm sin}(\Omega t_c)\frac{\pi}{2kd^2T}{\rm Re}[f_1(0)-f_2(0)],
\end{equation}
where we recall that the result is valid for the particular case where $\Omega T \simeq \pi$, that $f_\gamma = f_\gamma(0)$ here, and that we work at first order in the interaction.

\subsubsection*{Discussion}
\begin{figure}
\includegraphics[scale=0.22]{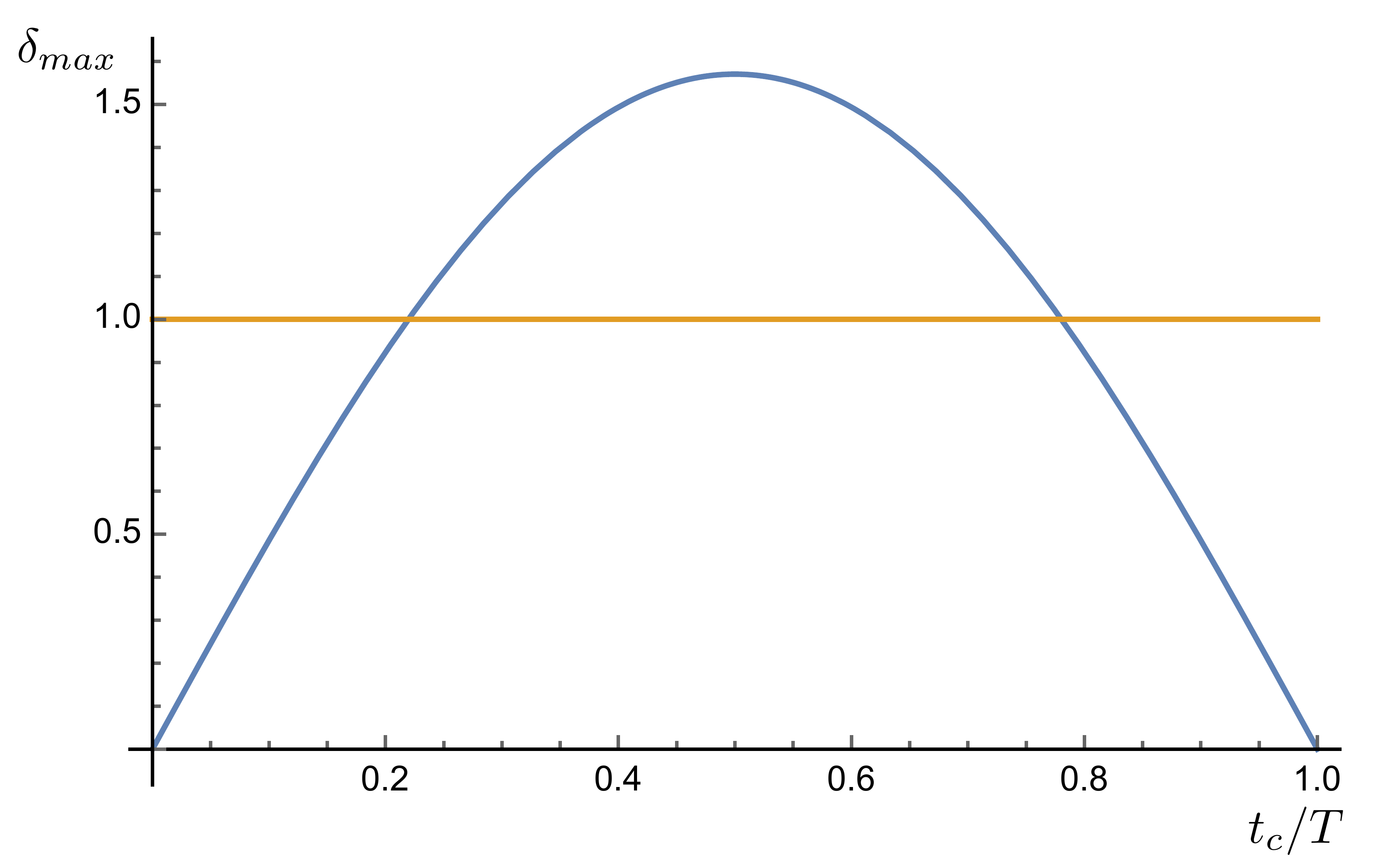}
\caption{Frequency shift (in units of ${\rm Re}[f_1-f_2]/(kd^2T)$) for Rabi (blue) and Ramsey (orange) spectroscopy as a function of the time $t_c$ at which scattering takes place.}
\label{fig_dmax}
\end{figure}

Equation (\ref{d_max_Rabi2}) is equivalent to the corresponding equation (\ref{d_max_Ramsey2}) for Ramsey spectroscopy, up to a factor $\frac{\pi}{2}{\rm sin}(\Omega t_c)$. This is consistent with the physical picture of the scattering provoking a transient shift of the atomic resonance frequency $\omega_{12}$. In the Rabi case, the effect of that shift on the spectroscopy depends on the time $t_c$ at which the collision occurs, contrary to the Ramsey sequence. As already mentioned, this can also be understood using the image of fictitious spin on a Bloch sphere or the sensitivity function formalism described e.g. in \cite{Santarelli1998}. Figure \ref{fig_dmax} shows the frequency shift for Rabi and Ramsey spectroscopy as a function of the time of the scattering $t_c$. For the Rabi case the shift is zero at $t_c=0$ and $t_c=T$ and maximum at $t_c=T/2$ as expected from the physical picture.

An interesting result is that when averaged over $T \simeq \pi/\Omega$ the Rabi frequency shift $\langle \delta_{max}(t_c) \rangle_T$ is identical to the Ramsey one i.e. $\left\langle \frac{\pi}{2}{\rm sin}(\Omega t_c)\right\rangle_T = 1$, as can be easily checked. In practice that means that {\it on average} the scattering frequency shift for Ramsey or Rabi spectroscopy is the same. However, the fluctuations around that average will be larger for Rabi spectroscopy, which  suggests one possible strategy for detecting scattering from light background particles, as discussed briefly in section \ref{sec:DM_strategies}.

\subsection{Summing over scatterers} \label{sec:sum_scat}

In all of the above we have considered a single scattering event during the Ramsey (Rabi) time $T$. The generalisation to several scattering events is straightforward when assuming that the difference Re$[f_1-f_2]$ is the same (or with non-vanishing average value) for the different scattering events.  This point is explained in detail in \cite{Alonso2017} for different models. In the following we simply assume that this quantity is the same for all the events. We further work in the limit $\Psi_{sc} \ll \Psi_{inc}$ which implies that for $N_{sc}$ events the final wave function is $\Psi_{out} = \Psi_{inc} + N_{sc}\Psi_{sc}$, i.e. if we neglect the scattering of the scattered wave. Then all the scattering shifts simply add and the total shift is the individual one multiplied by $N_{sc}$. As shown in Appendix \ref{app:A} we have $N_{sc}\simeq 2n\pi d^2vT$, where $n$ is the scatterer number density.

For Ramsey spectroscopy the total frequency shift is then obtained by multiplying (\ref{d_max_Ramsey2}) by $N_{sc}$ which gives
\begin{equation} \label{d_max_Ramsey2_n}
\delta_{max} \simeq \frac{2\pi n \hbar}{\mu}{\rm Re}[f_1(0)-f_2(0)].
\end{equation}
The resulting frequency shift is independent of the Ramsey time $T$ and the relative velocity $v$ between the atom and the scatterers (provided $f$ is independent of $v$). Both of these effects can be understood physically: For a single scatterer the larger the velocity the shorter the transient shift in $\omega_{12}$ relative to the Ramsey time $T$ i.e. the shift decreases with $T$ and with $v$ (hence $T$ and $k$ in the denominator of (\ref{d_max_Ramsey2})). But the larger $T$ and $v$ the larger the number of scattering events $N_{sc}$ and thus $T$ and $v$ cancel in the final result.

For Rabi spectroscopy the calculation is somewhat more involved and requires the distribution in time of the number of scattering events per unit time i.e. $dN(t)/dt$ to be used in (\ref{d_max_Rabi2}). The resulting frequency shift is
\begin{equation} \label{d_max_Rabi_n2}
\delta_{max} \simeq \frac{\pi^2 n \hbar}{\mu N_ {sc}}{\rm Re}[f_1-f_2] \int_0^T dt_c\frac{dN(t_c)}{dt_c}{\rm sin}(\Omega t_c) 
\end{equation}
for the special case where $\Omega T \simeq \pi$ as in equ. (\ref{d_max_Rabi2}).

For a uniform distribution with $dN(t)/dt=$~const. the resulting frequency shift is identical to (\ref{d_max_Ramsey2_n}) as already discussed at the end of section \ref{sec:Rabi}.

\section{Dark matter detection strategies} \label{sec:DM_strategies}

In this section we propose some strategies for detecting possible collisions between the atoms in clocks and light dark matter (DM) particles with $m_\chi \ll m_a$. The methods described here are by no means exhaustive, but may be useful for some particular DM models as, for example, discussed in the companion article \cite{Alonso2017}.

The different strategies are accompanied by very rough order of magnitude sensitivity estimates based on present day technology of Cs or Rb clocks (see e.g. \cite{Guena2012,Guena2014}). Depending on the particular DM model other clocks may be more adapted, so the orders of magnitude given here provide only some benchmark numbers without exhausting the full potential of atomic clock detection techniques.

The aim here is to provide rough numbers to evaluate the potential sensitivity of atomic clocks to any DM model, provided it allows the derivation of scattering amplitudes for interactions with atoms. If the resulting sensitivity to the model parameters turns out to be interesting a more detailed and specific analysis is required, that may then lead to a dedicated experiment.

\subsection{Measuring an offset or slow modulations} \label{sec:offset}

The average frequency of the clock transition will be shifted by an amount given by equation (\ref{d_max_Ramsey2_n}). However, to measure the shift one needs a reference frequency that is not affected by DM collisions. 

A first strategy for strong enough interactions is  shielding a reference clock that operates on the same atomic transition. Depending on the DM model under study such shielding could be provided by e.g. the Earth itself or the Earth's atmosphere.  
One possible configuration would consist of two identical clocks on opposite sides of the Earth compared via satellite links. As the Earth moves through galactic DM at $10^{-3}c$ it acts as a shield with more DM present on the ``forward" side. As a result, the measured frequency difference will oscillate at the Earth rotation frequency, a signal that could be searched for in the data. Possibly the search could be further refined and confirmed by comparing clocks on the surface and in deep underground mines.

If a large fraction of DM is absorbed in the atmosphere, then one should be able to observe a constant frequency difference between ground and space clocks operating on the same transition, like e.g. in the upcoming ACES (Atomic Clock Ensemble in Space) mission that will install a high performance Cs clock on board the International Space Station in 2020, together with high performance microwave and optical links that will allow frequency comparison to ground clocks \cite{Salomon2009}.

If the scattering amplitudes depend on the relative velocity of DM with respect to the atoms, one would also expect to observe modulations as the amplitude and direction of that velocity vary due to the rotation of the Earth and its orbital motion. If the interaction is spin dependent and DM is polarized then one would expect modulations as the orientation of the atomic spins (when using $m_F \neq 0$ states) varies with the Earth's rotation.

Finally, one can also try to compare with clocks where Re$[f_2-f_1]= 0.$ This is the natural option in the case of DM interacting with the spin of the atomic state. In this particular situation, the comparison of clocks operating at different value of $m_F $  would generate a modulated signal \cite{Alonso2017} (we elaborate more on this possibility below).

The standard clock transition used in $^{133}$Cs clocks is the $|F=3,m_F=0\rangle \leftrightarrow |F=4,m_F=0\rangle$ hyperfine transition of the $6S_{1/2}$ ground state at $\approx$\,9.2~GHz. In $^{87}$Rb clocks it is the $|F=1,m_F=0\rangle \leftrightarrow |F=2,m_F=0\rangle$ hyperfine transition of the $5S_{1/2}$ ground state at $\approx\,$6.8~GHz. In both cases, and for any of the above scenarios, a rough order of magnitude for the uncertainty in the determination of the frequency shift is $\Delta\delta \approx 10^{-5}$~rad/s \cite{Guena2012,Guena2014}. Inserting that into (\ref{d_max_Ramsey2_n}) one can expect a sensitivity of 
\begin{equation} \label{limit1}
\frac{\rho_\chi \hbar}{m_\chi^2}{\rm Re}[f_1(0)-f_2(0)] \approx \frac{\Delta\delta}{2\pi},
\end{equation}
where $\rho_\chi$ is the local density of the $\chi$ particles (0.4 GeV/cm$^3$ is expected if all of DM is made up of $\chi$ \cite{Read:2014qva}) and $\mu \simeq m_\chi$ the DM mass. Equation (\ref{limit1}) can be used to set constraints on the difference in scattering amplitudes for the two atomic states as a function of DM mass. Note that the constraint on the scattering amplitudes becomes more stringent for smaller DM masses, i.e. the sensitivity of clocks is best for small DM masses.

Many DM models involve spin dependence and may lead to a vanishing value of ${\rm Re}[f_1(0)-f_2(0)]$ for the $m_F=0$ states generally used in clocks and assumed in the estimate (\ref{limit1}) above. They may however be sensitive to transitions involving $m_F \neq 0$ Zeeman states and would lead to additional modulation effects (see above). Unfortunately the Zeeman states are first order sensitive to magnetic fields which strongly degrades the uncertainty in the frequency measurement. One way around that limitation is to use simultaneously two different atoms or isotopes in the same magnetic field, and form a combination that is first order insensitive to the Zeeman effect in spite of the $m_F \neq 0$ states involved. If that combination retains sensitivity to the DM model that is tested a precise measurement can be performed. As a particular example consider $^{133}$Cs and $^{87}$Rb that are measured simultaneously in the same magnetic environment in the dual fountain clock FO2 at SYRTE \cite{Guena2012,Guena2014}. Defining $\nu_i^{\rm Cs}$ as the frequency of the $|F=3,m_F=1\rangle \leftrightarrow |F=4,m_F=i\rangle$ Zeeman transition of $^{133}$Cs and similarly for $^{87}$Rb, the combination $K(\nu_1^{\rm Cs}-\nu_0^{\rm Cs})-(\nu_1^{\rm Rb}-\nu_0^{\rm Rb})$ is free of magnetic effects to leading order with $K\approx2$. Then we directly obtain a sensitivity of
\begin{equation} \label{limit1b}
\frac{\rho_\chi \hbar}{m_\chi^2}\left(K{\rm Re}[f_1-f_2]^{\rm Cs}-{\rm Re}[f_1-f_2]^{\rm Rb}\right) \approx \frac{\Delta\delta}{2\pi},
\end{equation}
which allows a meaningful search provided that the DM model does not imply a vanishing left hand side of (\ref{limit1b}). As shown in the companion paper \cite{Alonso2017} this is quite naturally the case in plausible DM models due to the different nuclear spin of the two atoms used. For the present order of magnitude estimates we will assume that the measurement of the Zeeman-free combination $2\nu_1^{\rm Cs}-\nu_1^{\rm Rb}$ can be carried out with the same uncertainty as the $m_F=0$ spectroscopy, i.e. $\Delta\delta \approx 10^{-5}$~rad/s in (\ref{limit1b}), particularly when searching for modulations at the rotation and orbital frequencies of the Earth as expected for spin dependent interactions.  

\subsection{Ramsey vs Rabi spectroscopy} \label{sec:RabivsRamsey}

As shown in section \ref{sec:main} the effect of scattering may lead to frequency fluctuations that are larger for Rabi spectroscopy than Ramsey spectroscopy (c.f. fig. \ref{fig_dmax}) when using the same atomic transition and the same interrogation time $T$. Therefore one could try and detect DM by comparing the clock stability of the same transition in the same clock but run on a Rabi or Ramsey sequence. For a single scattering event during the interrogation time $T$ and for a single atom the increase in the fluctuations when using Rabi interrogation  is given by the standard deviation of the frequency shift (\ref{d_max_Rabi_n2}) as can be also seen on figure \ref{fig_dmax}. It is
\begin{eqnarray} \label{sigma_Rabi}
\sigma &=& \frac{2\pi n \hbar}{\mu}{\rm Re}[f_1-f_2] \left( \frac{1}{T}\int_0^T d t_c\left( \frac{\pi}{2}{\rm sin}(\Omega t_c)-1\right)^2  \right)^{1/2} \nonumber \\
&=& \frac{2\pi n \hbar}{\mu}{\rm Re}[f_1-f_2] \left(\frac{\pi^2}{8}-1\right)^{1/2},
\end{eqnarray}
where $T\simeq \pi/\Omega$. When averaging over $N_a$ atoms and $N_{sc}$ scattering events the value of $\sigma$ is decreased by a factor $1/\sqrt{N_aN_{sc}}$.

The frequency fluctuations of the best Cs or Rb atomic clocks are limited by atomic shot noise at about $\sigma_a\approx 10^{-3}$~rad/s for a single interrogation cycle of the $m_F=0$ clock transition with $N_a\approx 5\times 10^6$ atoms \cite{Guena2012}. It is proportional to $1/T$ and to $1/\sqrt{N_a}$. Although this is achieved in Ramsey spectroscopy only, we will somewhat optimistically assume that similar stability is possible in Rabi spectroscopy in a dedicated experiment. The resulting sensitivity is then
\begin{equation} \label{limit2}
\frac{\rho_\chi \hbar}{m_\chi^2\sqrt{N_aN_{sc}}}{\rm Re}[f_1(0)-f_2(0)] \approx \frac{\sigma_a}{2\pi\left(\frac{\pi^2}{8}-1\right)^{1/2}}.
\end{equation}

A rough estimation of $N_{sc}$ is provided in Appendix \ref{app:A}. As $N_{sc}$ is proportional to $T$, and $\sigma_a$ is proportional to $1/T$, the sensitivity improves with $\sqrt{T}$. On the other hand $\sigma_a$ is proportional to $1/\sqrt{N_a}$ indicating that the sensitivity is independent of the total number of atoms $N_a$ used. 

For spin dependent $m_F\neq 0$ transitions the stability is in general limited by the magnetic field instability. E.g. for the $|F=3,m_F=1\rangle \leftrightarrow |F=4,m_F=1\rangle$ Cs transition, the instability is typically about three orders of magnitude worse than for the standard $m_F=0$ clock transition. However, one could operate the clock with much less atoms so that the dominant noise is again atomic shot noise. The overall sensitivity then remains the same as in the $m_F=0$ case as explained in the previous paragraph. Finally, one could of course use Zeeman-free combinations of different atoms as discussed in the previous section.

\subsection{Summary} \label{sec:summary_sens}

Table \ref{tab:summary} gives some numerical sensitivity estimates for the scenarios discussed above with particular assumptions as specified. We stress again that these numbers are only rough benchmarks for potential sensitivity that should allow evaluating, for any particular DM model, whether a more detailed analysis, possibly followed by a dedicated experiment, is worth pursuing.

\begin{table*} \label{tab:summary}
\caption{Numerical sensitivity estimates for the different detection strategies described above. We provide the minimum scattering amplitudes (or their combination) that could be detected, together with the assumptions made.}
\begin{ruledtabular}
\begin{tabular}{ccc}
Section & min. detectable $f$/m & Assumptions\\
\hline
\ref{sec:offset}  & ${\rm Re}[f_1(0)-f_2(0)] \geq 10^{-22} \left(\frac{\rho_\chi}{0.4 \,{\rm GeV/cm}^3}\right)(m_\chi/{\rm eV})^2$ & $m_F=0$, shielding or velocity dependence\\
\ref{sec:offset}  & ${\rm Re}[f_1(0)-f_2(0)] \geq \times 10^{-22} \left(\frac{\rho_\chi}{0.4 \,{\rm GeV/cm}^3}\right)(m_\chi/{\rm eV})^2$ & $m_F\neq 0$, Zeeman-free Rb-Cs combination, \vspace{-.12cm}\\
& & vel. dep. or polarized DM \\
\hline
\ref{sec:RabivsRamsey} & ${\rm Re}[f_1(0)-f_2(0)] \geq 10^{-14} \ \left(\frac{\rho_\chi}{0.4 \,{\rm GeV/cm}^3}\right)(m_\chi/{\rm eV})^{3/2}$ & Rabi vs. Ramsey, $m_\chi>10^4$ eV\\
\ref{sec:RabivsRamsey} & ${\rm Re}[f_1(0)-f_2(0)] \geq 10^{-10} \ \left(\frac{\rho_\chi}{0.4 \,{\rm GeV/cm}^3}\right)(m_\chi/{\rm eV})^{1/2}$ & Rabi vs. Ramsey, $m_\chi<10^4$ eV\\
\end{tabular}
\end{ruledtabular}
\label{SME_tab}
\end{table*}

\section{Conclusion}

We have presented a theoretical analysis of the sensitivity of atomic clocks to interaction with Dark Matter (DM) beyond the more usual massive scalar field models widely used in recent years \cite{Derevianko2014, Arvanitaki2015,Stadnik2015a, VanTilburg2015,Hees2016,Wcislo2016a, Roberts2017a,Hees2018}. More particularly we focus on DM whose mass is much less than the mass of the atoms in the clocks and density such that the occupation numbers are $< 1$ and hence cannot be treated as classical fields. We have analysed the scattering effects on the clock frequency, showing that such effects are first order in the differential forward scattering amplitudes $f_1(0)-f_2(0)$ of the involved atomic states. The effect has a simple physical interpretation as a transient differential shift of the atomic state energies by the DM-atom interaction potential as the DM particle flies past the atom.

As a consequence different detection strategies can be imagined, which we describe in section \ref{sec:DM_strategies}. We also give some very rough order of magnitude sensitivities to the differential scattering amplitudes as a function of DM mass and density in Tab. \ref{tab:summary}, that can be confronted to any DM model that can be expressed in terms of such differential scattering amplitudes. A number of such models are treated in much more detail in the companion paper \cite{Alonso2017}. As discussed in more detail in that paper, for certain models atomic co-magnetometers (e.g. \cite{Brown:2010dt, Allmendinger:2013eya}) can be treated in a very similar way as the clocks in this work, and may provide competitive bounds as well.

We hope that our work will inspire confronting different DM models to atomic clock (or co-magnetometer) data, with possibly dedicated or opportunistic experiments in the near future. Our results are also relevant for understanding the effects of other astrophysical backgrounds in atomic clocks, such as astrophysical neutrinos or gravitational waves. We hope to explore these ideas further in the future.

\acknowledgments
Helpful discussions with Kurt Gibble and Luigi de Sarlo are gratefully acknowledged.

\appendix

\section{Partial wave expansion} \label{app:partial}

We provide a partial wave expansion of the full detection probability (\ref{P2_3}). To do so we use
\begin{equation} \label{equ:f-partial}
f_\gamma(\theta) = \frac{1}{2ik}\sum\limits_{\ell=0}^\infty (2\ell+1)\left[e^{2i\delta_{\gamma\ell}} -1\right] P_\ell({\rm cos}\theta)
\end{equation}
where $\delta_{\gamma\ell}$ are the partial wave phases and $P_\ell(x)$ the Legendre polynomials. Then the integral in $F_{12}$ can be evaluated over all solid angles by making use of the orthogonality $\int_{-1}^1dx P_n(x)P_m(x) = 2\delta_{nm}/(2n+1)$ giving
\begin{equation}
F_{12} = \frac{\pi}{k^2}\sum\limits_{\ell=0}^\infty (2\ell+1)\left[e^{2i\delta_{1\ell}} -1\right]\left[e^{2i\delta_{2\ell}} -1\right],
\end{equation}
which when separated into real and imaginary parts is

\begin{eqnarray} \label{equ:F12-partial}
{\rm Re}[F_{12}] &=& \frac{4\pi}{k^2}\sum\limits_{\ell=0}^\infty (2\ell+1)\left({\rm sin}^2\delta_{1\ell}{\rm sin}^2\delta_{2\ell}\right. \nonumber \\
& & \hspace{2cm} \left. + {\rm sin}(2\delta_{1\ell}){\rm sin}(2\delta_{2\ell})/4 \right) \\
{\rm Im}[F_{12}] &=& \frac{4\pi}{k^2}\sum\limits_{\ell=0}^\infty (2\ell+1)\left({\rm sin}\delta_{1\ell}{\rm sin}\delta_{2\ell}{\rm sin}(\delta_{1\ell}-\delta_{2\ell}) \right). \nonumber
\end{eqnarray}
Substituting (\ref{equ:F12-partial}) into (\ref{P2_3}) and replacing the $f_\gamma(0)$  by (\ref{equ:f-partial}) one obtains after some simplification
\begin{eqnarray} \label{equ:P_2-partial}
P_2 &=& \frac{1}{2}(1+{\rm cos}\phi) \nonumber \\
&-& \frac{1}{2k^2d^2} \sum\limits_{\ell=0}^\infty (2\ell+1) {\rm sin}^2(\delta_{1\ell}-\delta_{2\ell}) {\rm cos}\phi \nonumber \\
&+& \frac{1}{4k^2d^2} \sum\limits_{\ell=0}^\infty (2\ell+1) {\rm sin}(2\delta_{1\ell}-2\delta_{2\ell}) {\rm sin}\phi.
\end{eqnarray}

When summing over all scatterers, i.e. multiplying (\ref{equ:P_2-partial}) by $N_{sc}\simeq 2n\pi d^2vT$ (c.f. section \ref{sec:sum_scat}) one obtains an expression that is equivalent to equ. (2) of \cite{Gibble2013}.

\section{Rough estimate of the number of scattering events} \label{app:A}

The number of scatterers in time $T$ crossing a circular area with radius $r_A$ is $N_{r_A} = n v T \pi r_A^2$.
The wavefunction $\Psi(t,{\bf x})$ describes the relative position of the two particles (see section \ref{sec:GenTheory}). We assume it is Gaussian with the corresponding probability distribution having a $1/e^2$ radius equal to $d$ (c.f. (\ref{equ:psi_inc})). The number of scattering events in time $T$ is then the probability of the relative position being $\leq r_a$ multiplied by $N_{r_A}$
\begin{equation}\label{equ:B4}
N_{sc} = nvT \pi r_A^2 \left(1 - e^{-\frac{r_A^2}{2d^2}} \right)^{-1},
\end{equation}
where $r_A$ is some characteristic length, typically the range of the interaction potential.

The relative wavefunction $|\Psi(t,{\bf x})|^2$ gives the probability of finding the two particles at a relative position ${\bf x}$ at time $t$. Therefore its radius $d$ is roughly equal to the radius of the larger one of the two single particle wavefunctions in the lab frame.

The minimum size of the atomic wavefunction is given by the uncertainty principle and the atomic temperature ($\mu$K, cm/s velocity spread) at roughly $d_{a} \approx 10^{-8}$ m.

The minimum size of the DM wavefunction is also given by the uncertainty principle and the virial velocity spread of DM in the Galaxy ($\Delta v \approx 10^{-3}c$) at roughly $d_{\chi} \approx 10^{-4}/(m_\chi/{\rm eV})$~m. So for DM masses $< 10^4 \ {\rm eV}$ it is $d_{\chi}$ that needs to be used in the expression for $N_{sc}$.
The DM number density is given by $n\approx 4\times 10^{14}/(m_\chi/{\rm eV})$~m$^{-3}$ for the  density of 0.4~GeV/cm$^3$.

Finally, the relevant radius $r_A$ is typically the range of the interaction potential. If   this is less than the value of $d$, i.e. $r_A \ll 10^{-8}$~m then equ. (\ref{equ:B4}) reduces to

\begin{equation}
N_{sc} \simeq 2\pi n d_a^2 vT \approx 10^5 \left(\frac{{\rm eV}}{m_\chi}\right)\left(\frac{\rho_\chi}{0.4 \,{\rm GeV/cm}^3}\right)
\end{equation}
for $m_\chi > 10^4\ {\rm eV}$, and 

\begin{equation}
N_{sc} \simeq 2\pi n d_\chi^2 vT \approx 10^{13} \left(\frac{{\rm eV}}{m_\chi}\right)^3\left(\frac{\rho_\chi}{0.4 \,{\rm GeV/cm}^3}\right)
\end{equation}
for $m_\chi \leq 10^4\ {\rm eV}$.

\bibliography{DM_colls}
\end{document}